\documentclass[conference, compsoc]{IEEEtran}
\IEEEoverridecommandlockouts
\usepackage[numbers]{natbib}
\usepackage[hyphens]{url}
\usepackage{booktabs}
\usepackage{amsmath,amssymb,amsfonts}
\usepackage{graphicx}
\usepackage{eurosym}
\usepackage{cuted}
\def\BibTeX{
    {\rm B\kern-.05em{\sc i\kern-.025em b}\kern-.08em
    T\kern-.1667em\lower.7ex\hbox{E}\kern-.125emX}
}

\begin{document}

\title{Identification of the Breach of Short-term Rental Regulations in Irish Rent Pressure Zones}

\author{
    \IEEEauthorblockN{Guowen Liu}
    \IEEEauthorblockA{
        \textit{School of Computer Science and Statistics} \\
        \textit{Trinity College Dublin} \\
        Dublin, Ireland \\
        liu.guowen@outlook.com
    }
    \and
    \IEEEauthorblockN{Inmaculada Arnedillo-S\'{a}nchez}
    \IEEEauthorblockA{
        \textit{School of Computer Science and Statistics} \\
        \textit{Trinity College Dublin} \\
        Dublin, Ireland \\
        macu.arnedillo@tcd.ie
    }
    \and
    \IEEEauthorblockN{Zhenshuo Chen}
    \IEEEauthorblockA{
        \textit{School of Computing} \\
        \textit{Dublin City University} \\
        Dublin, Ireland \\
        chenzs108@outlook.com
    }
}

\maketitle

\begin{abstract}
The housing crisis in Ireland has rapidly grown in recent years.
To make a more significant profit, many landlords are no longer renting out their houses under long-term tenancies but under short-term tenancies.
The shift from long-term to short-term rentals has harmed the supply of private housing rentals.
Regulating rentals in Rent Pressure Zones with the highest and rising rents is becoming a tricky issue.

In this paper, we develop a breach identifier to check short-term rentals located in Rent Pressure Zones with potential breaches only using publicly available data from Airbnb (an online marketplace focused on short-term home-stays).
First, we use a Residual Neural Network to filter out outdoor landscape photos that negatively impact identifying whether an owner has multiple rentals in a Rent Pressure Zone.
Second, a Siamese Neural Network is used to compare the similarity of indoor photos to determine if multiple rental posts correspond to the same residence.
Next, we use the Haversine algorithm to locate short-term rentals within a circle centered on the coordinate of a permit.
Short-term rentals with a permit will not be restricted.
Finally, we improve the occupancy estimation model combined with sentiment analysis, which may provide higher accuracy.

Because Airbnb does not disclose accurate house coordinates and occupancy data,
it is impossible to verify the accuracy of our breach identifier.
The accuracy of the occupancy estimator cannot be verified either.
It only provides an estimate within a reasonable range.
Users should be skeptical of short-term rentals that are flagged as possible breaches.
\end{abstract}

\begin{IEEEkeywords}
Housing Crisis, Short-term Rental, Irish Rent Pressure Zone, Image Recognition, Breach Identification
\end{IEEEkeywords}

\section{INTRODUCTION}
The housing crisis in Ireland has rapidly grown in recent years.
To make a more significant profit, many landlords are no longer renting out their houses under long-term tenancies but under short-term tenancies.
Under Irish regulations, \emph{Short-term Tenancies} are defined as those for less than 14 days.
According to Airdna \cite{airdna-rental-data} and Residential Tenancies Board \cite{residential-tenancies-board-report}, the average monthly revenue from short-term rentals is greater than the revenue from long-term rentals as Table.~\ref{tbl:rent-revenue},
which has resulted in the number of available long-term rentals continuing to decline while lucrative short-term rentals are becoming more popular.

\begin{table*}[htbp]
    \caption{The average monthly rents in the first quarter of 2022}
    \begin{center}
        \begin{tabular}{ccc}
            \toprule
            Area & Long-term Average Monthly Rent & Short-term Average Monthly Rent \\
            \midrule
            Dublin & \euro$2183$ & \euro$2622$ \\
            Galway & \euro$1413$ & \euro$2874$ \\
            Cork & \euro$1453$ & \euro$2097$ \\
            \bottomrule
        \end{tabular}
    \end{center}
    \label{tbl:rent-revenue}
\end{table*}

The shift from long-term to short-term rentals has harmed the supply of private housing rentals.
During the housing crisis, the number of houses used for tourism is higher than those used for private rentals in every county except Dublin.
According to Inside Airbnb \cite{inside-airbnb}, the number of rental properties on Airbnb\footnote{\url{https://www.airbnb.ie}} in the Irish capital alone has increased from around $1700$ in 2016 to more than $7064$ in 2022, $97.2\%$ of which are short-term rentals.
This is even more serious in those popular tourist counties.

As a ripple effect, rental prices have increased significantly.
According to Daft.ie \cite{daft.ie-report}, the rental market in Ireland has hit new highs as the number of available homes has fallen to historically low levels.
Rents in the second quarter of 2022 rose by an average of $12.6\%$ compared to the same period in 2021.
This is the largest increase since Daft.ie began reporting in 2006 and an all-time low in the number of homes available for rent, with the scarcity of available homes in Ireland now unprecedented.
The supply of rental properties has fallen by a staggering $97\%$ compared to 2009.
Only $716$ homes were available for rent nationwide on August 1, down from nearly $2500$ a year ago and a record low since 2006.

The government has approved measures to tighten regulation of the short-term rental market.
Under the new regulations, short-term rental platforms such as Airbnb and landlords will be fined if they advertise non-compliant properties from September 1, 2022 \cite{short-term-property-rental}.
The regulations target areas where the accommodation shortage is most acute, requiring Airbnb to obtain planning permission before advertising on their platforms in Rent Pressure Zones (see Section.~\ref{sct:rent-pressure-zone}).
In 2021, $286$ short-term rentals were advertised on Airbnb in Cork.
In the same year, Cork Council launched $94$ investigations and issued $68$ warning letters.
However, there are still $11$ counties that have not taken any action against these short-term rental owners on the popular website to check if they are in violation, making it a difficult issue to regulate short-term rentals in Rent Pressure Zones \cite{counties-clamp-down-rental}.
The Office of the Planning Regulator previously stated that regulating short-term rentals in Rent Pressure Zones is becoming tricky.
On August 2022, the Irish government has not yet provided an efficient way to investigate it, which motivated us to write this paper.

This paper demonstrates how to identify short-term rentals located in Rent Pressure Zones with potential violations using only publicly available data from Airbnb.
\begin{enumerate}
    \item
    We use a Residual Neural Network to identify and filter out outdoor landscape photos.
    These noisy photos negatively impact identifying whether an owner has multiple short-term rentals in a Rent Pressure Zone.

    \item
    A Siamese Neural Network is used to compare the similarity of images to determine if the interior regions of any two houses are highly similar.
    There are cases in Airbnb where the same home is posted multiple times or different rooms in the same home are posted multiple times.

    \item
    We use the Haversine algorithm to locate short-term rentals within a circle with a radius of $150$ meters, centered on the coordinate of a permit.
    Short-term rentals with a permit will not be restricted.

    \item
    We estimate the occupancy of a house by its user reviews combined with sentiment analysis.
\end{enumerate}

\section{BACKGROUND}
\subsection{Rent Pressure Zones}
\label{sct:rent-pressure-zone}
A \emph{Rent Pressure Zone} is a designated area in Ireland where rents cannot exceed the general inflation rate, as recorded by the Harmonized Index of Consumer Prices, or $2\%$ per year pro rata.
This applies to both new and existing tenancies (unless an exemption is being applied for).
Rent Pressure Zones are located in the parts of the country with the highest and rising rents and where households have the most difficulty finding affordable accommodation \cite{residential-tenancies}.
They are designed to moderate rent increases in these areas and create a stable and sustainable rental market, allowing owners and tenants to plan financially for their future.

For areas to be designated as Rent Pressure Zones, the following criteria are used:
\begin{itemize}
    \item
    The standardized average rent for the previous quarter must be higher than the appropriate reference standardized average rent for that quarter.

    \item
    Annual rental inflation in the area must have been $7\%$ or higher in four of the last six quarters.
\end{itemize}

\begin{figure}[htbp]
    \centerline{\includegraphics[width=0.9\linewidth]{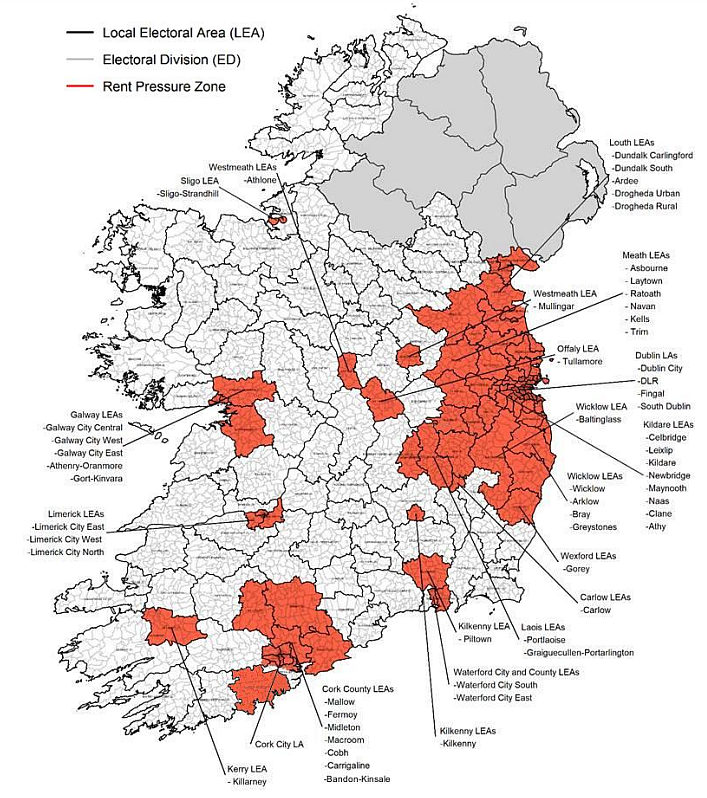}}
    \caption{The map of Irish Rent Pressure Zones \cite{housing-property-sector-chartpack}}
    \label{fig:rent-pressure-zone-map}
\end{figure}

Ireland now has three different standardized average rents, and the area's location determines the assessment of which method is used.
Dublin is now compared to the national standardized average rent.
Kildare, Meath and Wicklow are now compared to the national standardized rent, excluding Dublin.
The other areas are now compared to the standardized average rent outside the Greater Dublin Area.
On July 2022, there are $48$ areas set as Rent Pressure Zones in six local authorities, as shown in Fig.~\ref{fig:rent-pressure-zone-map}:
\begin{itemize}
    \item Dublin City Council
    \item South Dublin County Council
    \item Cork City Council
    \item Rathdown County Council
    \item Fingal County Council
    \item Kildare County Council
\end{itemize}

\subsection{Planning Permission}
According to \cite{planning-development-regulations}, if an owner rents out an entire house or one of its rooms on a short-term basis, they may need to apply for planning permission from the local authority to use it for short-term rentals.
The requirement for planning permission only applies to owners in Rent Pressure Zones.
Once planning permission is granted, they can use the entire principal residence for short-term rentals of more than $90$ days or rent out a second house for short-term rentals when they are not at home.

\subsection{New Regulation of Short-term Rentals}
The following rules are the new regulation for short-term rentals \cite{planning-development-regulations}:
\begin{itemize}
    \item
    Short-term letting is the letting of a house or apartment, or part of a house or apartment, for any period not exceeding $14$ days.

    \item
    Letting one or more rooms in the owner's principal residence in a private home will be allowed without restriction.

    \item
    Home sharing will continue to be permissible unrestricted and exempt from the new planning requirements.
    Where the $90$ day threshold is exceeded, change of use planning permission will be required.

    \item
    If the house or flat is not a principal residence, the $90$-day exemption does not apply, and planning permission is required.
\end{itemize}

In the following cases, planning permission is not required \cite{citizen-planning-permission}:
\begin{itemize}
    \item
    The property is not in a Rent Pressure Zone.

    \item
    The property is in a Rent Pressure Zone, but the rental period is $15$ days or more at a time.

    \item
    The property already has planning permission for tourism or short-term rental purposes.

    \item
    The property is used for corporate or executive rentals.

    \item
    The property is rented out under the rent-a-room scheme.

    \item
    One or more rooms are rented out in the owner's principal residence on a family-share basis for short periods.
    The owner must also occupy the home at the same time.

    \item
    The entire principal private residence is rented out for short-term visitors for less than $90$ days a year while the owner is temporarily away.
    The $90$ days do not have to be consecutive.

    \item
    The property is purpose-built student accommodation.
\end{itemize}

\section{DATA COLLECTION}
\subsection{Airbnb House and Owner Data}
\begin{figure}[htbp]
    \centerline{\includegraphics[width=\linewidth]{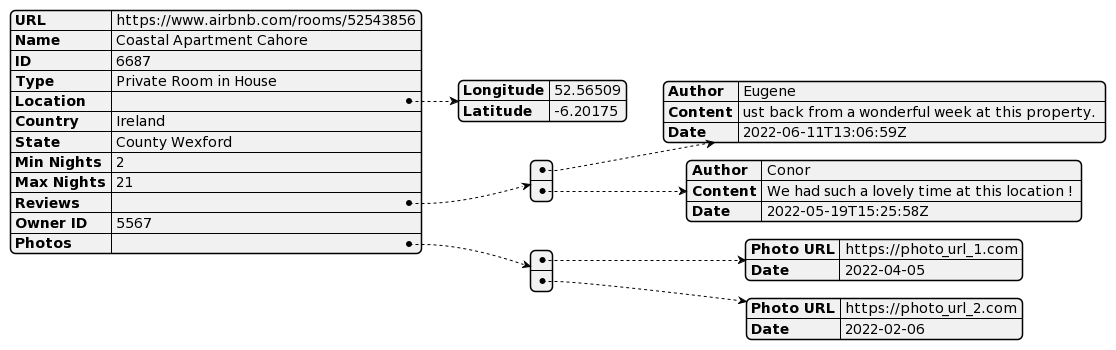}}
    \caption{An Airbnb house entry}
    \label{fig:airbnb-house-entry}
\end{figure}

\begin{figure}[htbp]
    \centerline{\includegraphics[width=0.6\linewidth]{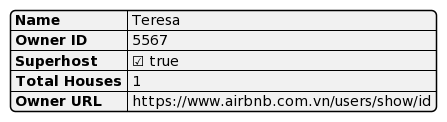}}
    \caption{An Airbnb owner entry}
    \label{fig:airbnb-owner-entry}
\end{figure}

The publicly accessible home data (Fig.~\ref{fig:airbnb-house-entry}) and owner data (Fig.~\ref{fig:airbnb-owner-entry}) were obtained using Airbnb APIs, including home type, reviews, home photos, the total number of an owner's homes, etc.
No private information was used.

An important issue is that those home locations were randomized by Airbnb.
In detail, the public geographic coordinate of a house in Airbnb data is a random location within a circle with a radius of $150$ meters, centered on its actual geographic coordinate.
So even if two rooms come from the same building, it is impossible to distinguish them directly.

\subsection{Geographic Information of Rent Pressure Zones}
\begin{figure}[htbp]
    \centerline{\includegraphics[width=\linewidth]{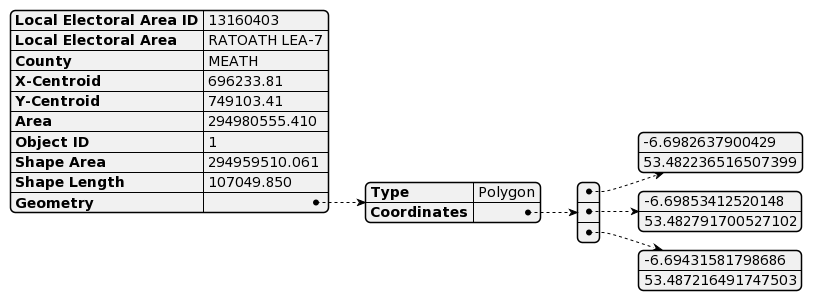}}
    \caption{A local electoral area entry}
    \label{fig:local-electoral-area-entry}
\end{figure}

The Local Electoral Area Dataset \cite{local-electoral-area-dataset} contains geographical features, as shown in Fig.~\ref{fig:local-electoral-area-entry}, which helps identify whether a house is located within a Rent Pressure Zone.

\subsection{National Planning Application Dataset}
\begin{figure}[htbp]
    \centerline{\includegraphics[width=\linewidth]{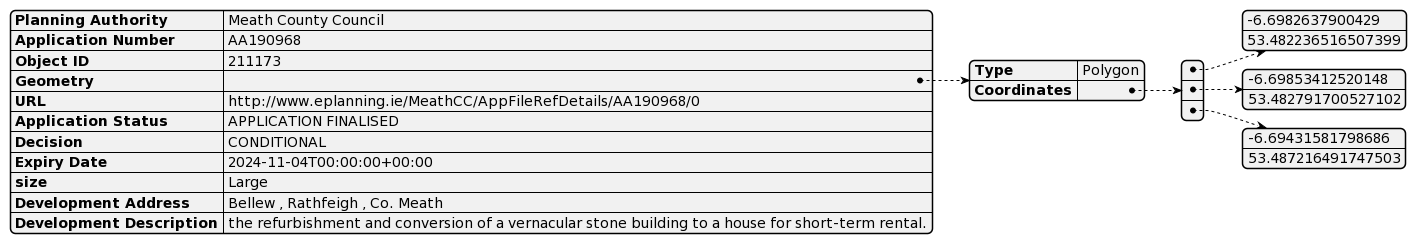}}
    \caption{A national planning application entry}
    \label{fig:national-planning-application-entry}
\end{figure}

The National Planning Application Dataset \cite{national-planning-application-dataset} provides details on all planning applications in Ireland and contains polygon coordinates of housing applications and descriptions of permission applications, as shown in Fig.~\ref{fig:national-planning-application-entry}.
This dataset was first filtered to obtain all entries related to short-term rental applications.

\section{DATA PREPROCESSING}
\subsection{Locating the Centroid of House}
\begin{figure}[htbp]
    \centerline{\includegraphics[width=0.5\linewidth]{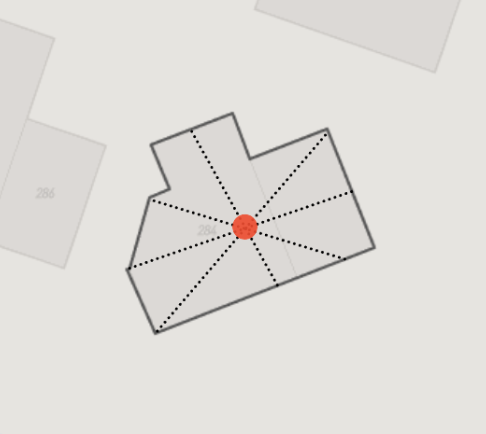}}
    \caption{The centroid of a house}
    \label{fig:house-centroid}
\end{figure}

The National Planning Application Dataset uses polygon coordinates to record the geographic location of a house.
The centroid coordinate was calculated and used to uniquely identify a house's location during this stage, as shown in Fig.~\ref{fig:house-centroid}.

\subsection{Filtering Houses in Rent Pressure Zones}
The PNPOLY algorithm filtered out houses not located in a Rent Pressure Zone.
This algorithm was proposed by \cite{pnpoly-algorithm} to solve the problem of determining whether a coordinate is inside a polygon for a geographic information management system.

\section{METHODOLOGY}
The breach identifier contains three components:
\begin{itemize}
    \item
    A \emph{Principal Residence Identifier} containing two components:
    \begin{itemize}
        \item
        An image classifier for indoor and outdoor.
        It deletes noisy outdoor photos before image similarity detection.
        \item
        An image similarity detector used to determine whether photos from different posts come from the same residence or room.
    \end{itemize}

    \item
    A \emph{Permit Finder}.
    Under the rules, it is necessary to determine whether a house has planning permission if it is not an owner's principal residence or its cumulative rental period within one year is more than $90$ days.
    The permit finder checks if a house is within the scope of a valid short-term rental permit.

    \item
    An \emph{Occupancy Estimator}.
    The rules require that when an owner entirely rents out their principal residence, it cannot exceed $90$ days.
    Due to trade secrecy issues, Airbnb does not disclose the occupancy time of a home, so the only way to estimate it is by the number of reviews.
\end{itemize}

\subsection{Principal Residence Identifier}
\begin{figure}[htbp]
    \centerline{\includegraphics[width=\linewidth]{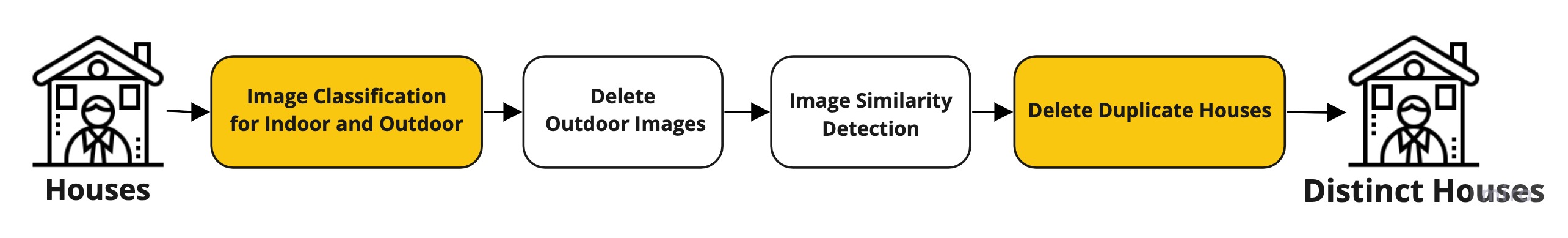}}
    \caption{The principal residence identifier}
    \label{fig:principal-residence-identifier}
\end{figure}

The principal residence identifier is based on the rules for short-term rentals within Rent Pressure Zones.
If a short-term rented house is not an owner's principal residence, they need to obtain the relevant planning permission to rent it out legally.
The flow chart is shown in Fig.~\ref{fig:principal-residence-identifier}, where an owner's all house photos are filtered through an image classifier based on a Residual Neural Network to remove outdoor photos,
then a Siamese Neural Network is used to compare the similarity of indoor photos to remove duplicate houses.
Unlike a model that learns to classify inputs, a Siamese Neural Network learns how to differentiate between two inputs by learning their similarity \cite{siamese-neural-network}.
It shows better performance in situations involving the discovery of similarity or relationship between two comparable things.

\subsubsection{Image Classification for Indoor and Outdoor}
\begin{figure}[htbp]
    \centerline{\includegraphics[width=\linewidth]{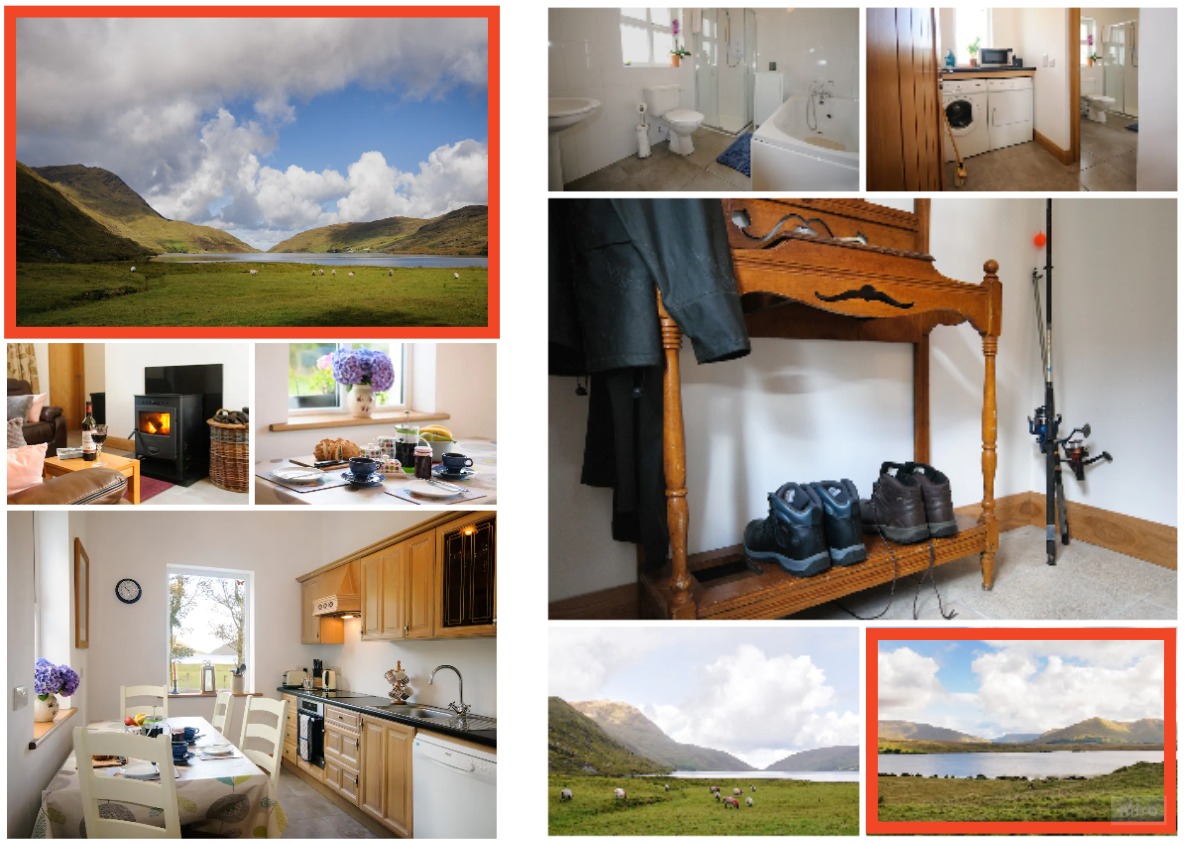}}
    \caption{Two posts have the same outdoor photo}
    \label{fig:outdoor-photos}
\end{figure}

Some posts on Airbnb contain many outdoor photos, such as beautiful landscapes, because owners want to use them to attract more visitors.
These noisy data can affect the subsequent identification of whether multiple short-term rentals correspond to the same house posted by an owner.
For example, two posts contain the same landscape photos in Fig.~\ref{fig:outdoor-photos}, but come from different houses.
So it is required to remove these noisy images.
The image classifier uses a Residual Neural Network ResNet-18 \cite{resnet18} to classify whether an image is indoor or outdoor and delete outdoor photos.

\subsubsection{Image Similarity Detection}
\begin{figure}[htbp]
    \centerline{\includegraphics[width=\linewidth]{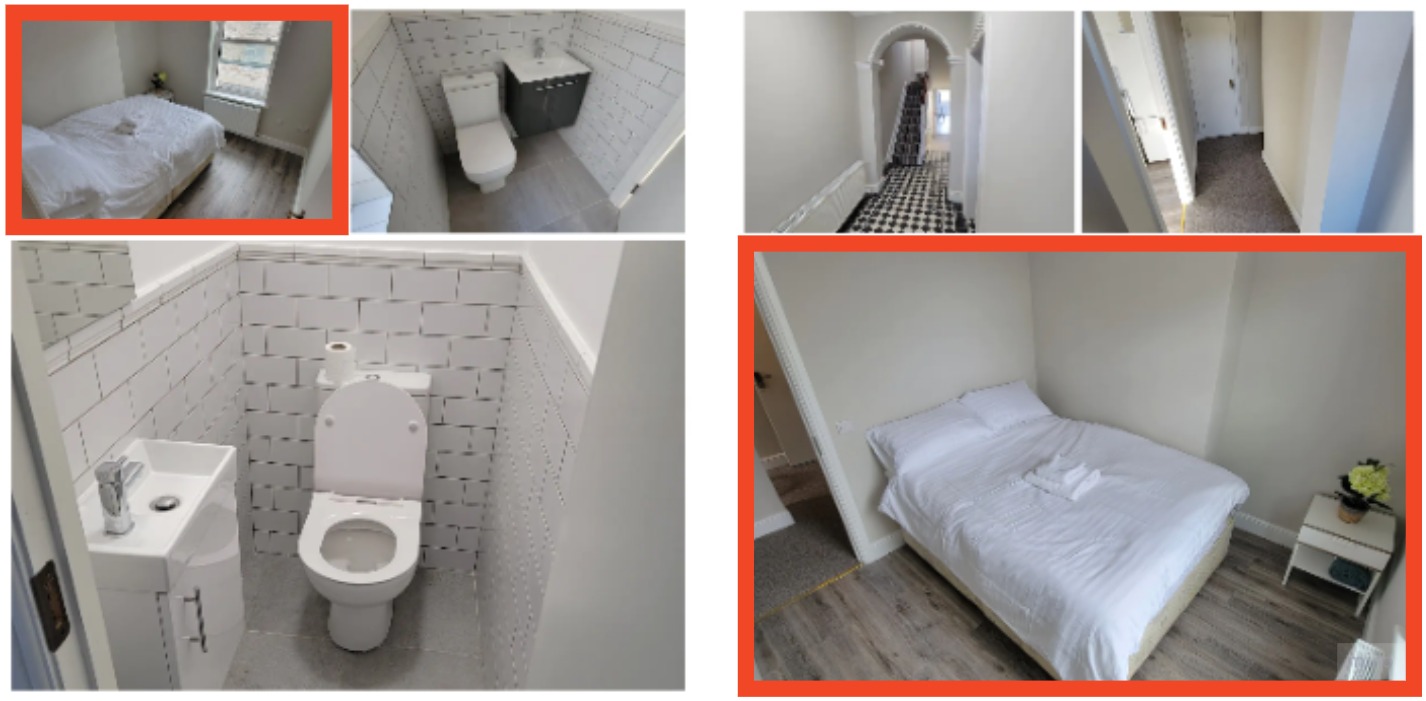}}
    \caption{Two posts correspond to the same room}
    \label{fig:same-room-photos}
\end{figure}

There are two legitimate scenarios in which the breach identifier may incorrectly assume that an owner has multiple homes in a Rent Pressure Zone, which is not allowed.
\begin{itemize}
    \item
    Owners can rent out multiple rooms in their principal residence without restriction.
    They can post two or more rooms in their principal residence on Airbnb.

    \item
    Owners may post the same room multiple times on Airbnb to give their home more exposure.
    In Fig.~\ref{fig:same-room-photos}, these two posts correspond to the same room, but the photos were taken from different angles.
\end{itemize}

\begin{figure}[htbp]
    \centerline{\includegraphics[width=\linewidth]{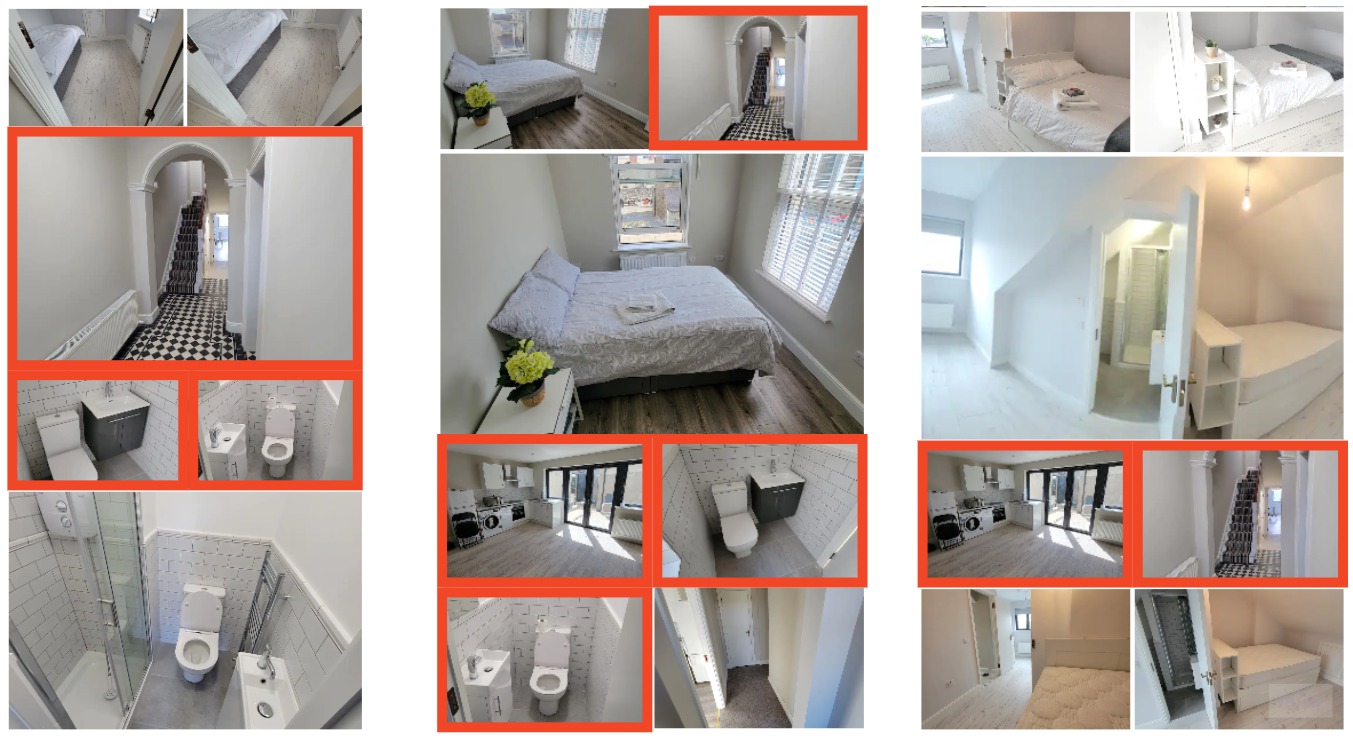}}
    \caption{Three posts correspond to the same house}
    \label{fig:same-house-photos}
\end{figure}

To address these two issues, we should check photos from different posts to see whether they come from the same principal residence or the same room.
If a room is posted multiple times, there must be many similar or the same photos.
If multiple rooms are from the same residence, their photos probably contain identical common areas such as the kitchen, hallway and bathroom.
For example, three posts in Fig.~\ref{fig:same-house-photos} are from different rooms in the same house.
They have identical common areas with a high degree of similarity.

In most cases, two images will not be fully identical even if they have the same scene or object due to differences in size, scale, rotation and skew.
So the comparison cannot be made directly using pixel-to-pixel methods.
The problem shifted from identifying pixel similarity to object similarity.

The \verb|sentence-transformers| library was used in this project to provide a way to compute a dense vector representation of an image and find identical images \cite{sentence-bert}.
It encodes all images into vector space and finds high-density regions corresponding to areas where the images are similar.
For example, when using image search, a photo is entered and converted into a set of vectors.
\verb|sentence-transformers| compares it with other images in the database and finds the image with the highest similarity (the closest distance).
When two photos from one post have greater than $95\%$ similarity to the photos from another post, or one photo has $100\%$ similarity to another, two posts are considered to be associated with the same room.
So the breach identifier will not assume its owner has violated the rules.

\subsection{Permit Finder}
\begin{figure}[htbp]
    \centerline{\includegraphics[width=0.5\linewidth]{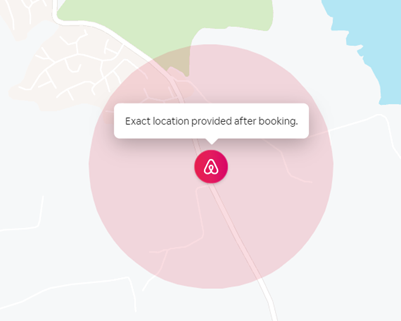}}
    \caption{The anonymous geographic coordinate of a house}
    \label{fig:anonymous-house-coordinate}
\end{figure}

As the actual geographic coordinate of a house has been anonymized by Airbnb, replaced with a random location within a circle with a radius of $150$ meters as Fig.~\ref{fig:anonymous-house-coordinate},
it is impossible to know whether a permit is available for a short-term rental house by conventional coordinate comparison.
We use the Haversine algorithm \cite{haversine-algorithm} to determine whether a house is within a circle with a radius of $150$ meters, centered on the location of any known permit.
By using the longitude and latitude of two locations, the Haversine algorithm can calculate their distances on the surface of the Earth.
Because of the anonymization issue, when there are multiple houses around a permit, they will all be considered to have the same available permit.

\subsection{Occupancy Estimator}
The occupancy estimator is based on the rule that, in some cases, short-term rentals are not allowed to accumulate more than $90$ days within a year.
One of the biggest issues with Airbnb is whether owners rent residential properties as hotels permanently rather than occasionally sharing.
On Airbnb, the occupancy time of a house is unpublished data. However, it can be estimated by the number of user reviews.

In \cite{sentiment-forecast}, \citeauthor{sentiment-forecast} found that the sentiment of a hotel's reviews affected its occupancy.
Inside Airbnb uses an occupancy estimation model named the \emph{San Francisco} model \cite{san-francisco-model} to estimate how often an Airbnb house is being rented out (Equation.~\ref{eq:san-francisco-model}) and approximate its income.
The review rate in Equation.~\ref{eq:san-francisco-model} can be different numbers.
Inside Airbnb set it to $50\%$, meaning half of the tenants write reviews about houses.
For the average number of nights of stay, Airbnb reported that it was $5.5$ nights in San Francisco.
If the average number of nights in a city is unknown, then three nights should be used.
Lastly, the minimum number of nights of stay can be set for a house by its owner.

\begin{strip}
    \begin{equation}
        \text{Occupancy} = \frac{\text{Number of Reviews}}{\text{Review Rate}} \times max(\text{Avg Nights},\, \text{Min Nights})
        \label{eq:san-francisco-model}
    \end{equation}
\end{strip}

In this project, an improvement for the San Francisco model was proposed.
Instead of a fixed review conversion rate, the improved model uses a rate of $50\%$ as a base and has a sentiment bias of $10\%$ increment or reduction depending on the sentiment score of house reviews.
Better reviews lead to more occupancy time.
The formula is Equation.~\ref{eq:sentimental-san-francisco-model},
where the average number of nights of stay for guests is $4.6$, taken from public Airbnb data,
and the minimum number of nights of stay can be set for a house by its owner.
Finally, the review rate is set as $50\%$.

\begin{strip}
    \begin{equation}
        \text{Occupancy} = \frac{\text{Number of Reviews}}{\text{Review Rate} + \text{Sentiment Bias}} \times max(\text{Avg Nights},\, \text{Min Nights})
        \label{eq:sentimental-san-francisco-model}
    \end{equation}
\end{strip}

\begin{figure}[htbp]
    \centerline{\includegraphics[width=\linewidth]{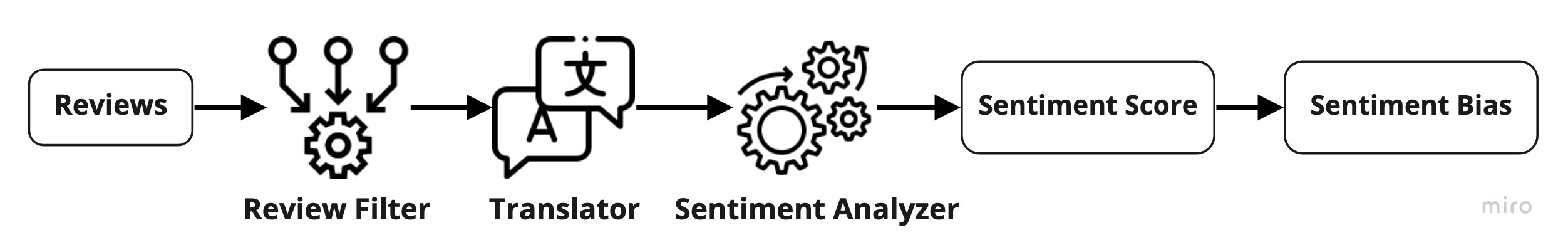}}
    \caption{The calculation of sentiment bias}
    \label{fig:sentiment-bias}
\end{figure}

The flow chart of sentiment bias calculation is shown in Fig.~\ref{fig:sentiment-bias}.
\verb|TextBlob| is introduced, which is a Python library for natural language processing \cite{textblob}.
It mainly uses pre-trained built-in classifiers to perform sentiment analysis and can obtain a corresponding sentiment score for a review.

\begin{enumerate}
    \item
    The reviews of a house are filtered to keep only the reviews within one past year.

    \item
    Reviews may be written in non-English languages, and DeepL API\footnote{\url{https://www.deepl.com/pro-api}} is used to translate those reviews to English automatically.

    \item
    The \verb|TextBlob| analyzer calculates an average sentiment score for all reviews.
    A sentiment score is between $[-1,\, 1]$ where $-1$ defines extremely negative emotion and $1$ defines extremely positive emotion.

    \item
    With the sentiment score, a sentiment bias can be produced using Equation.~\ref{eq:sentiment-bias}.
\end{enumerate}

\begin{equation}
    \text{Sentiment Bias} = \text{Sentiment Score} \times 0.1
    \label{eq:sentiment-bias}
\end{equation}

\section{EVALUATION}
There were $1713$ short-term rentals located in Rent Pressure Zones in the collected data.
$845$ of them were identified as having potential breaches, and $21$ principal houses with occupancy greater than $70$ days were about to breach the rule.
The overall breach rate was $50.67\%$
The reality can be worse, as a permit only works for one house in the actual situation.
However, the anonymous coordinates of houses may cause several houses to be simultaneously located within an available permit, resulting in a situation where one permit is used for multiple houses.

Because Airbnb does not disclose accurate house coordinates and occupancy data,
individuals cannot verify the accuracy of the breach identifier.
The accuracy of the occupancy estimator cannot be verified either.
It only provides an estimate within a reasonable range.
Tenants should be skeptical of short-term rentals that are flagged as possible breaches.

\section{CONCLUSION}
In summary, this project achieved the following contributions and benefits to the current regulation of the Irish short-term rental market:
\begin{itemize}
    \item
    It provided a viable method to identify potential short-term rentals that may be in violation and their owners when no accurate data is available.

    \item
    It provided a more efficient way to monitor the short-term rental market rather than relying on traditional user reporting.

    \item
    It provided an image identifier that can determine whether multiple rental posts correspond to the same room or residence by calculating the similarity of room photos.
    This model may be applied to other areas.
    For instance, some second-hand goods websites may prohibit users from posting duplicate items.

    \item
    It improved the occupancy estimation model combined with sentiment analysis, which may provide higher accuracy.
    The improved model may also be used for hotel occupancy estimation, etc.
\end{itemize}

\bibliographystyle{IEEEtranN}
\bibliography{rental-breach-identification}

\end{document}